\documentclass{article}
\usepackage{spconf,amsmath,graphicx}
\usepackage{amssymb,fixmath}
\usepackage{subfig}
\usepackage{float}
\usepackage{cite}
\usepackage[dvips]{epsfig}
\usepackage{multirow}
\usepackage{makecell}
\usepackage{multicol}
\usepackage{tabularx}
\usepackage{color}

\DeclareMathOperator{\E}{\mathbb{E}}
\DeclareMathOperator{\bx}{\mathbold{x}}
\DeclareMathOperator{\bz}{\mathbold{z}}
\DeclareMathOperator{\bh}{\mathbold{h}}
\DeclareMathOperator{\bxhat}{\mathbold{\hat{x}}}

\DeclareMathOperator{\bW}{\mathbold{W}}

\PassOptionsToPackage{hyphens}{url}\usepackage{hyperref}
\title{Improved Parallel WaveGAN Vocoder with Perceptually Weighted Spectrogram Loss}
\name{Eunwoo Song$^1$, Ryuichi Yamamoto$^2$, Min-Jae Hwang$^3$, Jin-Seob Kim$^1$, Ohsung Kwon$^1$, Jae-Min Kim$^1$}
\address{
  {$^1$NAVER Corp., Seongnam, Korea}\\
  {$^2$LINE Corp., Tokyo, Japan}\\ 
  $^3$Search Solutions Inc., Seongnam, Korea
  }
\begin{document}
%\fontsize{10}{11.5}\selectfont
%\ninept
%
\maketitle
%

% green red blue black magenta
\newcommand{\eunwooedit}[1]{\textcolor{black}{#1}}
\newcommand{\ryuichiedit}[1]{\textcolor{black}{#1}}
\newcommand{\minjaeedit}[1]{\textcolor{black}{#1}}

\newcommand{\feunwooedit}[1]{\textcolor{black}{#1}}
\newcommand{\fryuichiedit}[1]{\textcolor{black}{#1}}
\newcommand{\fminjaeedit}[1]{\textcolor{black}{#1}}

\newcommand{\reunwooedit}[1]{\textcolor{black}{#1}}
\newcommand{\rryuichiedit}[1]{\textcolor{black}{#1}}
\newcommand{\rminjaeedit}[1]{\textcolor{black}{#1}}

\begin{abstract}
    \reunwooedit{This paper proposes} a spectral-domain perceptual weighting technique for Parallel WaveGAN-based text-to-speech (TTS) systems. 
    The recently proposed Parallel WaveGAN vocoder successfully generates waveform sequences using a fast non-autoregressive WaveNet model.
    By employing \reunwooedit{multi-resolution short-time Fourier transform (MR-STFT) criteria} with \reunwooedit{a generative adversarial network}, the light-weight convolutional networks can be effectively trained without any distillation process. 
    To further improve the vocoding performance, we propose the application of frequency-dependent weighting to the MR-STFT loss function.
    The \minjaeedit{proposed} method penalizes perceptually-sensitive errors in the frequency domain\minjaeedit{; thus,} \reunwooedit{the model is optimized toward reducing auditory noise in the synthesized speech.}
    \reunwooedit{Subjective} listening test results demonstrate that our proposed method achieves \reunwooedit{4.21 and 4.26} \eunwooedit{TTS} mean opinion scores for female and male Korean speakers, respectively.
    \end{abstract}
    

\section{Introduction}
    Generative models for raw speech waveforms have significantly improved the quality of neural text-to-speech (TTS) systems \cite{ze2013statistical, Oord2016WaveNetAG}.
    Specifically, autoregressive generative models such as \textit{WaveNet} have successfully replaced the role of traditional parametric vocoders \cite{Oord2016WaveNetAG, tamamori2017speaker, hwang2018lp, song2019excitnet}.
    Non-autoregressive versions, including \textit{Parallel WaveNet}, provide a fast waveform generation method based on a teacher-student framework \ryuichiedit{\cite{Oord2018ParallelWF,ping2018clarinet}}.
    In this method, \minjaeedit{the model is trained \reunwooedit{using} a} probability density distillation \minjaeedit{method \reunwooedit{in which}} the knowledge of an autoregressive teacher WaveNet \minjaeedit{is transferred to} an inverse autoregressive flow student model \cite{Kingma2016ImprovingVI}. 
    
    \feunwooedit{In our previous work, we introduced generative adversarial network (GAN) training methods to the Parallel WaveNet framework \cite{Yamamoto2019}, and proposed \textit{Parallel WaveGAN} by combining the adversarial training with multi-resolution short-time Fourier transform (MR-STFT) criteria \cite{yamamoto2020parallel, yamamoto2020parallel2}.}
    %\rminjaeedit{Recently, }\reunwooedit{Yamamoto et al. \cite{Yamamoto2019} introduced} generative adversarial network (GAN) training methods \reunwooedit{to} the Parallel WaveNet framework, and proposed \textit{Parallel WaveGAN} by combining \rryuichiedit{the} adversarial training with multi-resolution short-time Fourier transform (MR-STFT) criteria \cite{yamamoto2020parallel}.
    Although it is possible to train GAN-based non-autoregressive models \reunwooedit{by only} using adversarial loss function \cite{kumar2019melgan}, \reunwooedit{employing the MR-STFT loss function has been proven to be advantageous for increasing the training efficiency \fryuichiedit{\cite{yang2020multi, yamamoto2020parallel, yang2020vocgan}.}}
    Furthermore, \reunwooedit{because} the Parallel WaveGAN only trains a WaveNet model without \minjaeedit{any} density distillation, the entire training process becomes much easier than \reunwooedit{it is in the} conventional methods, and the model can produce natural sounding speech waveforms with just a small number of parameters.
    
    To further enhance the performance of the Parallel WaveGAN, this paper proposes a spectral-domain perceptual weighting method for \eunwooedit{optimizing} the MR-STFT criteria. 
    A frequency-dependent masking filter is designed to penalize errors near the spectral valleys, which are perceptually-sensitive to the human ear \cite{schroeder1979optimizing}.
    By applying this filter to the STFT loss function calculations in the training step, \reunwooedit{the network is guided to reduce the noise component in those regions.}
    \reunwooedit{Consequently, the proposed model generates a more natural voice in comparison to the original Parallel WaveGAN.}
    Our contributions can be summarized as follows: 
    \begin{itemize}
    \item We propose a perceptually weighted MR-STFT loss function alongside a conventional adversarial training method. 
    This approach improves the \reunwooedit{quality} of the synthesized speech in the Parallel WaveGAN-based neural TTS system. 
    % RY 11/09: reduce the use of "can"
    \item \reunwooedit{Because} the proposed method does not change the network architecture, it \fryuichiedit{maintains} the \reunwooedit{small number of parameters found in the original Parallel WaveGAN’s and retain} its fast inference speed.
    %\item \reunwooedit{Because} the proposed method does not change the network architecture, it can maintain the \reunwooedit{small number of parameters found in the original Parallel WaveGAN’s and retain} its fast inference speed.
    In particular, the system can generate a 24 kHz speech waveform 50.57 times faster than real-time in a single GPU environment with 1.83 M parameters. 
    \item Consequently, our method achieved mean opinion score (MOS) results of \reunwooedit{4.21 and 4.26} for female and male Korean speakers, respectively, in the neural TTS systems.
    \end{itemize}

\section{Related work}
\label{sec:related work}

    The idea of using STFT-based loss functions is not new.
    In \reunwooedit{their} study of spectrogram inversion, Sercan et al. \cite{arik2019fast} first proposed \textit{spectral convergence} and \textit{log-scale STFT magnitude} losses, \feunwooedit{and our previous work proposed combing these in a multi-resolution form \cite{Yamamoto2019}}.
    %\reunwooedit{and Yamamoto et al. \cite{Yamamoto2019}} proposed \reunwooedit{combing} these in a multi-resolution form.
    
    \reunwooedit{Moreover}, perceptual noise-shaping filters have significantly improved the quality of synthesized speech in autoregressive WaveNet frameworks \cite{tachibana2018investigation}.
    Based on the characteristics of the human auditory system, \eunwooedit{an external noise-shaping} filter is \reunwooedit{designed to reduce perceptually-sensitive noise in the spectral valley regions.}
    \reunwooedit{This filter acts as a pre-processor in the training step; thus, the WaveNet learns the distribution of \textit{noise-shaped residual signal}.}
    \reunwooedit{In the synthesis step, by applying its inverse filter to the WaveNet's output, the enhanced speech can be reconstructed.}

    However, \rminjaeedit{it has been shown that} the filter's effectiveness does not work for the non-autoregressive generation models, including WaveGlow \cite{Okamoto2019} and the Parallel WaveGAN.
    \reunwooedit{One possible reason for this might be that the characteristics of the noise-shaped residual signal are difficult for the non-autoregressive model to capture without previous time-step information.}
    \reunwooedit{To address this problem, the proposed system applies a frequency-dependent mask} to the process of calculating the STFT loss functions.
    \reunwooedit{As this method does not change the target speech's distribution, the non-autoregressive WaveNet can be stably optimized, while significantly reducing the auditory noise components.}

    \section{Parallel WaveGAN}
    \label{sec:method}

    The Parallel WaveGAN jointly trains a non-causal WaveNet generator, $G$, and a convolutional neural network (CNN) discriminator, $D$, to generate a time-domain speech waveform from the corresponding input acoustic parameters. 
    Specifically, the generator learns a distribution of realistic waveforms by trying to deceive the discriminator into recognizing the generated samples as real. 
    The process is performed by minimizing the generator losses as follows:
    \begin{equation}
        L_{\mathrm{G}}(G, D) = L_{\mathrm{mr\_stft}}(G) + \lambda_{\mathrm{adv}}L_{\mathrm{adv}}(G, D), \label{eq:gloss}
    \end{equation}
    where $L_{\mathrm{mr\_stft}}(G)$ denotes an MR-STFT loss, which will be discussed in the next section; $L_{\mathrm{adv}}(G, D)$ denotes an adversarial loss; $\lambda_{\mathrm{adv}}$ denotes the hyperparameter balancing the two loss functions.
    \ryuichiedit{\eunwooedit{The} adversarial loss is designed based on least-squares GANs \cite{mao2017least, tian2018generative,Bollepalli2017,pascual2017segan}, as follows}:
    \begin{equation}
        L_{\mathrm{adv}}(G, D) = \E_{\bz \sim \ryuichiedit{p_{\bz}}}\left[(1 - D(G(\bz, \bh)))^2\right],
        \label{eq:adv}
    \end{equation}    
    \ryuichiedit{where $\bz$, $p_{\bz}$, and $\bh$ denote the input noise, a Gaussian distribution $N(\mathbf{0}, \mathbf{I})$, \eunwooedit{and the conditional} acoustic parameters, respectively.} 

    The discriminator is trained to correctly classify the generated sample as \textit{fake} while classifying the ground truth as \textit{real} using the following optimization criterion:
    \begin{equation}
        L_{\mathrm{D}}(G,D) = \E_{\bx \sim p_{\mathrm{data}}}[(1 - D(\bx))^2] + \ryuichiedit{\E_{\bz \sim p_{\mathrm{z}}}\left[D(G(\bz, \bh))^2\right]}, \label{eq:dloss}
    \end{equation}
    where $\bx$ and $p_{\mathrm{data}}$ denote the target speech waveform and its distribution, respectively.

    \subsection{\reunwooedit{Conventional MR-STFT} loss}

    To guarantee the stability of the adversarial training method described above, it is crucial to incorporate an MR-STFT loss function into the generator's optimization process \cite{yamamoto2020parallel}.
    The MR-STFT loss function in equation~(\ref{eq:gloss}) is defined in terms of the number of STFT losses, $M$, as follows:
    \begin{equation}
        L_{\mathrm{mr\_stft}}(G) = \frac{1}{M}\sum_{m=1}^{M}L^{(m)}_{\mathrm{stft}}(G), \label{eq:multispec}
    \end{equation}
    where $L^{(m)}_{\mathrm{stft}}(G)$ denotes the $m^{th}$ STFT loss defined as follows:
    \begin{equation}
        L_{\mathrm{stft}}(G) = \E_{\fminjaeedit{\bx \sim p_{data}, \bxhat \sim p_{G}}} \left[L_{\mathrm{sc}}(\bx,\bxhat) + L_{\mathrm{mag}}(\bx,\bxhat)\right], \label{eq:stftloss}
        % L_{\mathrm{stft}}(G) = \E_{\bz \sim \ryuichiedit{p_{\bz}}, \bx \sim p_{data}} \left[L_{\mathrm{sc}}(\bx,\bxhat) + L_{\mathrm{mag}}(\bx,\bxhat)\right], \label{eq:stftloss}
    \end{equation}
    where $\bxhat$ denotes the generated sample \fminjaeedit{drawn by probability distribution of generator, $p_G$}; $L_{\mathrm{sc}}$ and $L_{\mathrm{mag}}$ denote \textit{spectral convergence} and \reunwooedit{\textit{log STFT magnitude} losses, respectively, which are defined as follows} \cite{arik2019fast}:
    % where $\bxhat$ denotes the generated sample (i.e. \ryuichiedit{$G(\bz, \bh)$}) \fminjaeedit{drawn by  }; $L_{\mathrm{sc}}$ and $L_{\mathrm{mag}}$ denote \textit{spectral convergence} and \reunwooedit{\textit{log STFT magnitude} losses, respectively, which are defined as follows} \cite{arik2019fast}:
    %

    \begin{equation}
    \rryuichiedit{
         L_{\mathrm{sc}}(\bx,\bxhat) = \frac{\sqrt{\sum_{t,f} (|\mathbf{X}_{t,f}|  - |\hat{\mathbf{X}}_{t,f}|)^{2}}}{\sqrt{\sum_{t,f} |\mathbf{X}_{t,f}|^2}},
    }
    \end{equation}
    \begin{equation}
    \rryuichiedit{
    \label{eq:mag}
         L_{\mathrm{mag}}(\bx,\bxhat) = \frac{\sum_{t,f} | \mathrm{log}|\mathbf{X}_{t,f}| - \mathrm{log}|\hat{\mathbf{X}}_{t,f}||}{T \cdot N} ,
    }
    \end{equation}
    \rryuichiedit{where $|\mathbf{X}_{t,f}|$ and $|\hat{\mathbf{X}}_{t,f}|$ denote the $f^{th}$ STFT magnitude of $\bx$ and $\bxhat$ at the time \rryuichiedit{frame} $t$, \reunwooedit{respectively}; $T$ and $N$ denote the number of frames and the number of frequency bins, respectively.}
    
\subsection{\reunwooedit{Proposed} perceptual weighting for MR-STFT loss}
\label{ssec:ch3-3}

%%%%%%%%%%%%%%%%%%% Fig: loss %%%%%%%%%%%%%%%%%%%%%%%%%%%%%%
	\begin{figure}[!t]
	\begin{minipage}[t]{.32\linewidth}
	\centerline{\epsfig{figure=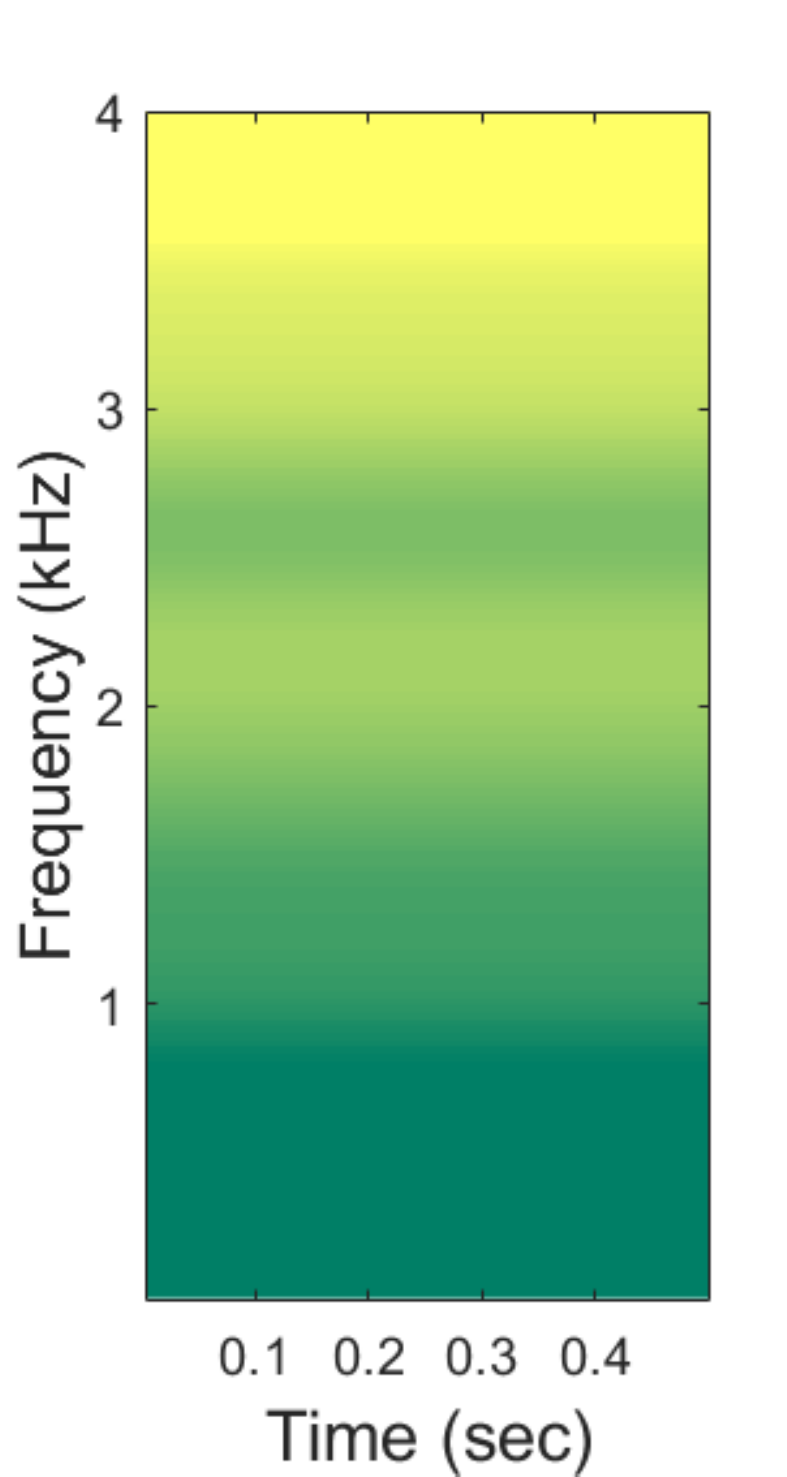,width=28mm}}
	%\vspace*{-1pt}	
	\centerline{(a)}  \medskip
	\end{minipage}
	\begin{minipage}[t]{.32\linewidth}
	\centerline{\epsfig{figure=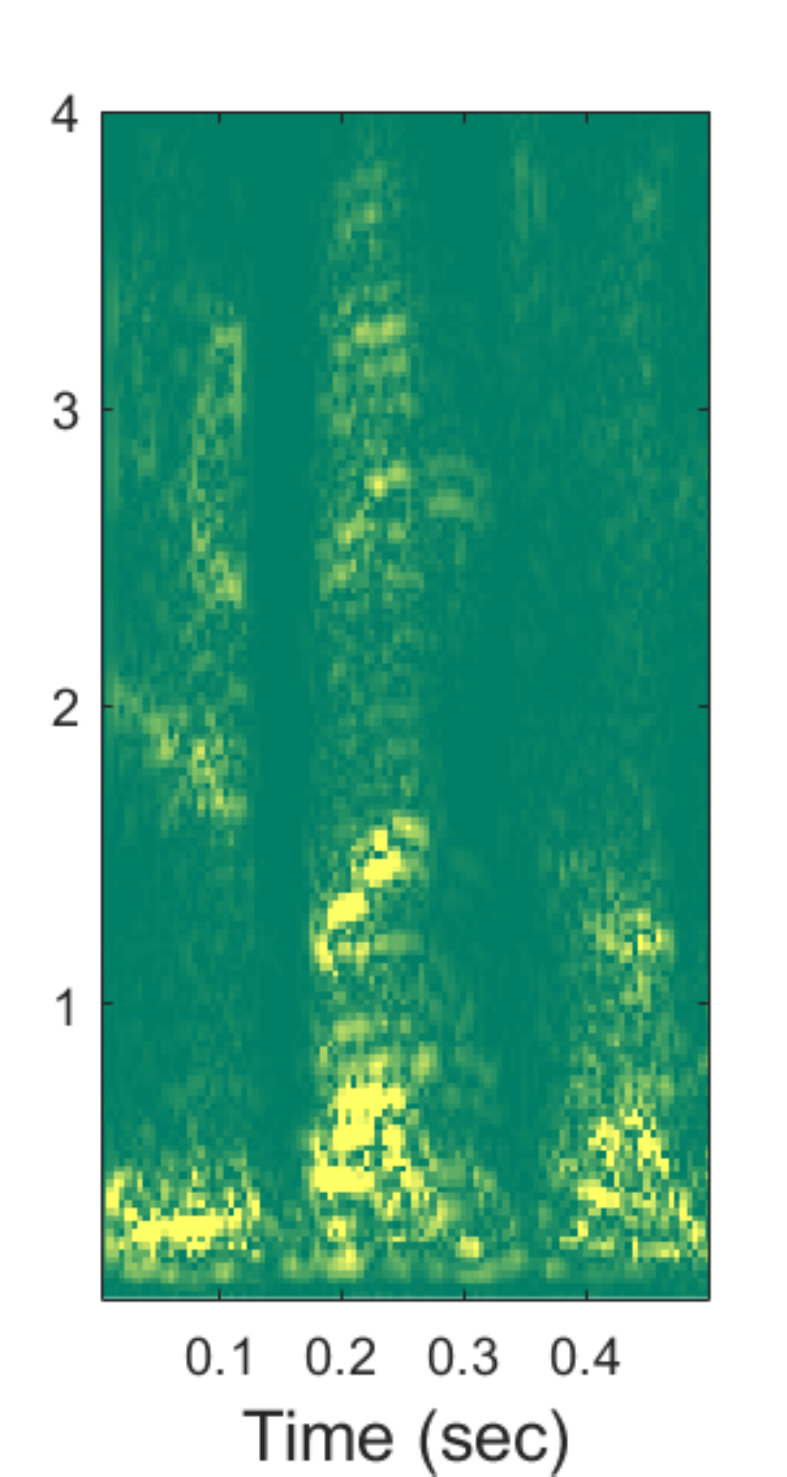,width=28mm}}
	%\vspace*{-1pt}	
	\centerline{(b)}  \medskip
	\end{minipage}	
	\begin{minipage}[t]{.32\linewidth}
	\centerline{\epsfig{figure=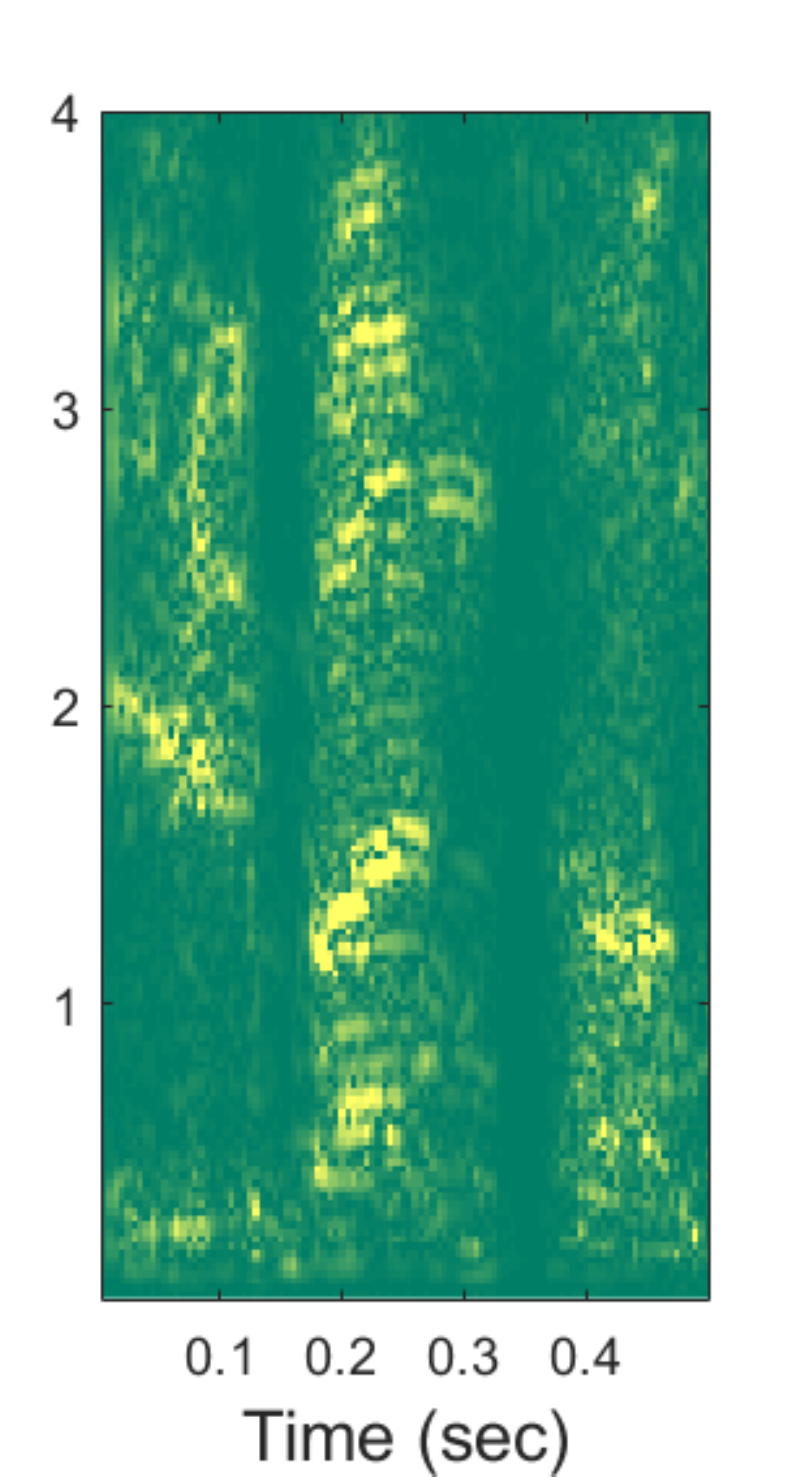,width=28mm}}
	%\vspace*{-1pt}	
	\centerline{(c)}  \medskip
	\end{minipage}	
	%\vspace*{-6pt} 
	\caption{Magnitude \ryuichiedit{distance} \eunwooedit{(MD)} obtained when calculating the spectral convergence: \eunwooedit{(a) The \rryuichiedit{weight matrix} of spectral mask, (b) the MD before applying the mask (conventional method), and (c) the MD after applying the mask (proposed method).}}	
	%\vspace*{-12pt} 
	\label{fig:loss}
	\end{figure}
%%%%%%%%%%%%%%%%%%% Fig: loss %%%%%%%%%%%%%%%%%%%%%%%%%%%%%%

    To further enhance the performance of the Parallel WaveGAN, this paper proposes \fminjaeedit{to} apply a spectral-domain perceptual \reunwooedit{masking} filter to the MR-STFT loss criteria as follows:
    \begin{equation}
    \rryuichiedit{
    \label{eq:opt_sc}
        L^{w}_{\mathrm{sc}}(\bx,\bxhat) = \frac{\sqrt{\sum_{t,f}(\mathbf{W}_{t,f} (|\mathbf{X}_{t,f}| - |\hat{\mathbf{X}}_{t,f}|))^2}}{\sqrt{\sum_{t,f}|\mathbf{X}_{t,f}|^2}},
    }
    \end{equation}
    \begin{equation}
    \rryuichiedit{    
    \label{eq:opt_mag}
        L^{w}_{\mathrm{mag}}(\bx,\bxhat) = \frac{\sum_{t,f} | \mathrm{log}\mathbf{W}_{t,f} (\mathrm{log}|\mathbf{X}_{t,f}
        | - \mathrm{log}|\hat{\mathbf{X}}_{t,f}|) |}{T \cdot N},
    }
    \end{equation}
    \rryuichiedit{where $\mathbf{W}_{t,f}$ denotes a weight coefficient of the spectral mask. The weight matrix $\mathbf{W}$ is constructed by repeating a time-invariant frequency masking filter along the time axis, whose transfer function is defined as follows:}
    \begin{equation}
    \label{eq:az}
         \bW(z) = 1 - \sum_{k=1}^{p}{\tilde{\alpha}_{k}z^{-k}},
    \end{equation}
    where $\tilde{\alpha}_{k}$ denotes the $k^{th}$ linear prediction (LP) coefficient with the order $p$, obtained by averaging all spectra extracted from the training data.
    % RY 11/09: applied grammarly (minor gramatical corrections!)
    \feunwooedit{
    As shown in Fig.~\ref{fig:loss}a, the weight matrix of the spectral mask is designed to represent \fryuichiedit{the} global characteristics of the spectral formant structure. 
    This enables an emphasis \fryuichiedit{on} losses at the frequency regions of the spectral valleys, which are more sensitive to the human ear. 
    When calculating the STFT loss (Fig.~\ref{fig:loss}b), this filter is used to penalize losses in those regions (Fig.~\ref{fig:loss}c). 
    As a result, the training process can guide the model to further reduce the perceptual noise in the synthesized speech\footnote{
    Although the log-scale STFT-magnitude loss in equation~(\ref{eq:mag}) was designed to fit small amplitude components \cite{arik2019fast}, our preliminary experiments verified that applying the masking filter to this loss was also beneficial to synthetic quality.}.
    }
    %\eunwooedit{As shown in Fig.~\ref{fig:loss}a, this} masking filter is designed to represent global characteristics of the spectral formant structure which enables an emphasis of losses at the frequency regions of the spectral valleys. 
    %\eunwooedit{Because} the human auditory system is more sensitive to noise components in these regions than at the spectral peaks \cite{schroeder1979optimizing}, by heavily penalizing losses near the valleys \eunwooedit{(Fig.~\ref{fig:loss}c)}, the training process can guide the model to further reduce the estimation errors in that region\footnote{
    %Although the log-scale STFT-magnitude loss in equation~(\ref{eq:mag}) was designed to fit small amplitude components \cite{arik2019fast}, our preliminary experiments verified that applying the masking filter to this loss was also beneficial to synthetic quality.}.
    
\begin{figure}[!t]
    \centerline{\epsfig{figure=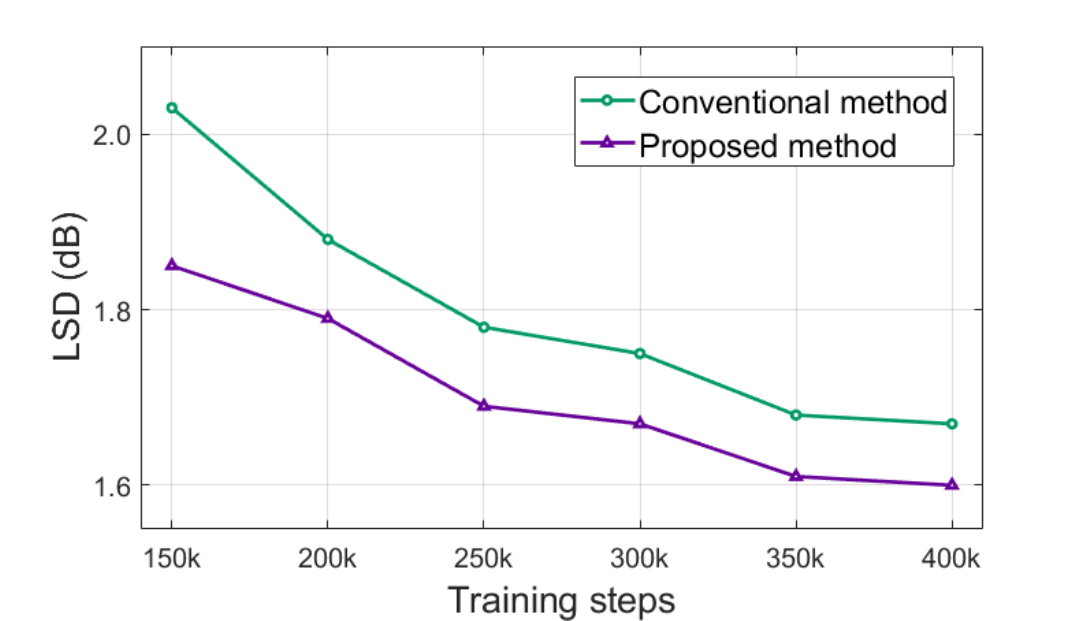,width=68mm}}
%    \vspace*{-6pt}  
    \caption{Log-spectral distance (LSD; dB) between the original and generated speech \eunwooedit{signals}}
%    \vspace*{-10pt}  
    \label{fig:lsd}
    \end{figure}

    The merits of the proposed method are presented in Fig.~\ref{fig:lsd} which shows the log-spectral distance between the original and generated speech \eunwooedit{signals}.
    The proposed perceptual weighting of MR-STFT losses enables an accurate estimation of speech spectra, and it is therefore expected that it will provide more accurate training and generation results, to be discussed further in the following section.

%%%%%%%%%%%%%%%%%%% Fig: lsd %%%%%%%%%%%%%%%%%%%%%%%%%%%%%%  
%%%%%%%%%%%%%%%%%%% Fig: lsd %%%%%%%%%%%%%%%%%%%%%%%%%%%%%%   

    \begin{figure*}[!t]
    \centerline{\epsfig{figure=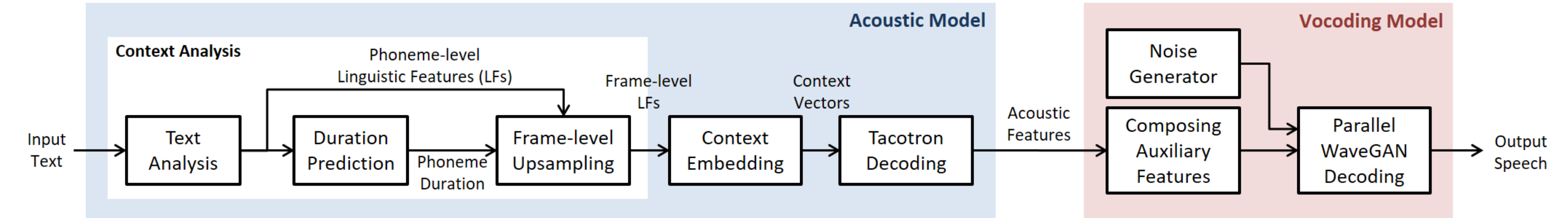,width=175mm}}
%    \vspace*{-6pt}  
    \caption{Block diagram of the TTS framework.}
%    \vspace*{-10pt}  
    \label{fig:tts}
    \end{figure*}
%%%%%%%%%%%%%%%%%%% Fig: am %%%%%%%%%%%%%%%%%%%%%%%%%%%%%%    

    %%%%%%%%%%%%%%% Table: Dataset %%%%%%%%%%%%%%%%%%%%%%%%
	\begin{table}[!t]   
	\begin{center}         
	\caption{Utterances in speech sets by Korean male (KRM) and Korean female (KRF) speakers (SPK).}  
%	\vspace*{-4pt}
	\label{table:numUtt}
	{\small        
	\begin{tabular}{>{\centering}m{.20\linewidth}ccc}
	\Xhline{2\arrayrulewidth}
	SPK			& Training  & validation 	& Testing \\
			\hline \hline
	KRF 	& 5,085 (5.5 h)		& 360 (0.4 h) 			& 180 (0.2 h)	\\
			%\hline
	KRM		& 5,382 (7.4 h)		& 290 (0.4 h)			& 140 (0.2 h)	\\
			\Xhline{2\arrayrulewidth}
	\end{tabular}}          
	\end{center}         
    %\vspace*{-8pt}
	\end{table}
    %%%%%%%%%%%%%%% Table: Dataset %%%%%%%%%%%%%%%%%%%%%%%%
    
    %%%%%%%%%%%%%%% Table: model %%%%%%%%%%%%%%%%%%%%%%%%
	\begin{table*}[!t]   
	\begin{center}         
	\caption{Vocoding model details, including size and inference speed: Note that inference speed, ${k}$, indicates that a system was able to generate waveforms ${k}$ times faster than real-time. This evaluation was conducted on a server with a single NVIDIA Tesla V100 GPU.}  
%	\vspace*{-4pt}
	\label{table:model}
	\scalebox{0.95}{
	{\small        
	\begin{tabular}{>{\centering}m{.10\linewidth}lccccccc}
	\Xhline{2\arrayrulewidth}
    \multirow{2}*{System} & \multirow{2}*{{~~~~~~~~~~~~~Model}} & \eunwooedit{MR-STFT} & {Perceptual} & {Noise} & {Number of} & {Model} & {Inference} \\
     & & \eunwooedit{loss} & {weighting} & {shaping} & {layers} & {size} & {speed} \\
    \hline \hline
    Baseline 1 & WaveNet & - & - & - & 24 & 3.71 M & 0.34$\times10^{-2}$ \\
    Baseline 2 & WaveNet + NS & - & - & Yes & 24 & 3.81 M & 0.34$\times10^{-2}$ \\
    Baseline 3 & Parallel WaveGAN & Yes & - & - & 30 & 1.83 M & 50.57  \\
    Baseline 4 & Parallel WaveGAN + NS & Yes & - & Yes & 30 & 1.83 M & 47.70  \\
    Proposal & Parallel WaveGAN + PW & Yes & Yes & - & 30 & 1.83 M & 50.57  \\			\Xhline{2\arrayrulewidth}
	\end{tabular}}    
	}
	\end{center}         
    %\vspace*{-12pt}
	\end{table*}
    %%%%%%%%%%%%%%% Table: model %%%%%%%%%%%%%%%%%%%%%%%%

\section{Experiments}
	\label{sec:experiment}
	\subsection{Experimental setup}
	\subsubsection{Database}
	The experiments used two phonetically and prosodically rich speech corpora recorded by Korean male and female professional speakers. 
	The speech signals were sampled at 24 kHz, and each sample was quantized by 16 bits. 
	Table~\ref{table:numUtt} shows the number of utterances in each set. 
	The acoustic features were extracted using an improved time-frequency trajectory excitation vocoder at \minjaeedit{the} analysis intervals of 5 ms \cite{song2017effective}, and these features included 40-dimensional line spectral frequencies (LSFs), fundamental frequency, energy, voicing flag, a 32-dimensional slowly evolving waveform, and a 4-dimensional rapidly evolving waveform, all of which constituted a 79-dimensional feature vector.
	
	\subsubsection{Acoustic model}
	Although there are many state-of-the-art acoustic architectures available \cite{wangIS2017tacotron, Shen2018NaturalTS, li2019close}, we used a Tacotron model with phoneme alignment \cite{okamoto2019tacotron, song2020neural} for its fast and stable generation and competitive synthesis quality. 
	The left section of Fig.~\ref{fig:tts} presents the acoustic model which consists of three sub-modules, namely, context analysis, context embedding, and Tacotron decoding. 
	
	In the context analysis module, \reunwooedit{a grapheme-to-phoneme converter was applied to the input text by the Korean standard pronunciation grammar, and then} phoneme-level feature vectors were extracted \reunwooedit{by the internal context information-labeling program}.
	These were composed of 330 binary features for categorical linguistic contexts and 24 features for numerical linguistic contexts. 
	By inputting those linguistic features, the corresponding phoneme duration was estimated through three fully connected (FC) layers with 1,024, 512, 256 units followed by a unidirectional long short-term memory (LSTM) network with 128 memory blocks.
	Based on this estimated duration, the phoneme-level linguistic features were then up-sampled to frame-level by adding the two numerical vectors of phoneme duration and relative position. 
	
	In the context embedding module, the linguistic features are transformed into high-level context vectors. 
	The module in this experiment consisted of three convolution layers with 10$\times$1 kernels and 512 channels per layer, a bi-directional LSTM network with 512 memory blocks, and an FC layer with 512 units. 
	
	To generate the output acoustic features, we used a Tacotron 2 decoder network \cite{Shen2018NaturalTS}. 
	First, the previously generated acoustic features were fed into two FC layers with 256 units (i.e. the PreNet), and those features and the vectors from the context embedding module were then passed through two uni-directional LSTM layers with 1,024 memory blocks followed by two projection layers. 
	Finally, to improve generation accuracy, five convolution layers with 5$\times$1 kernels and 512 channels per layer were used as a post-processing network (i.e. the PostNet) to add the residual elements of the generated acoustic features. 
    
    Before training, the input and output features were normalized to have zero mean and unit variance.
    The weights were initialized using \textit{Xavier} initialization \cite{xavier2010init} and \textit{Adam} optimization was used \cite{diederik2014adam}. 
    The learning rate was scheduled to be decayed from 0.001 to 0.0001 via a decaying rate of 0.33 per \eunwooedit{100 K} steps.

    \subsubsection{Vocoding model}

    Table~\ref{table:model} presents details of the vocoding models including their size and inference speed. 
    % RY 11/09: how about removing footnote? 
    As baseline systems, we used two \fryuichiedit{autoregressive WaveNet} vocoders, namely, a plain WaveNet (Baseline 1) \cite{tamamori2017speaker} and a WaveNet with noise-shaping (NS) method (Baseline 2) \cite{tachibana2018investigation}. \fryuichiedit{We adopted continuous Gaussian output distributions for both the baseline systems \cite{ping2018clarinet}, instead of using the categorical distributions.}
    %As baseline systems, we used two autoregressive Gaussian WaveNet vocoders\footnote{\feunwooedit{We adopted continuous Gaussian output distributions for both the baseline systems \cite{ping2018clarinet}, instead of using the categorical distribution.}}, namely, a plain WaveNet (Baseline 1) \cite{tamamori2017speaker} and a WaveNet with noise-shaping (NS) method (Baseline 2) \cite{tachibana2018investigation}.     
    These two approaches used the same network architecture but differed in \fryuichiedit{the} target output; the plain WaveNet system was \ryuichiedit{designed} to \ryuichiedit{predict} speech \eunwooedit{signals, whereas the latter \fryuichiedit{method}} was \ryuichiedit{designed} to \ryuichiedit{predict} the \ryuichiedit{noise-\eunwooedit{shaped}} residual signals. 
    \eunwooedit{Note that a} time-invariant \eunwooedit{noise-shaping} filter was \ryuichiedit{obtained} by averaging all spectra extracted from the training data.
    This \eunwooedit{external} \ryuichiedit{filter} was used to extract the residual signal before the training process, and its inverse filter was applied to reconstruct the speech signal in the synthesis step.
    
    The WaveNet systems consisted of 24 layers of dilated residual convolution blocks with four dilation cycles. 
    There were 128 residual and skip channels, and the filter size was set to three. 
    The model was trained for 1 M steps with a RAdam optimizer. The learning rate was set to 0.001, and this was reduced by half every 200 K steps. The minibatch size was set to eight, and each audio clip was set to 12 K time samples (0.5 seconds). 
    
    The experiment involved three Parallel WaveGAN systems, namely, the plain Parallel WaveGAN (Baseline 3) \cite{yamamoto2020parallel}, a Parallel WaveGAN with the same noise-shaping method as before (Baseline 4), and the proposed method with the perceptually \eunwooedit{weighted \rryuichiedit{(PW)}} criteria (Proposal).
    All had the same network architecture consisting of 30 dilated residual convolution block layers with three exponentially increasing dilation cycles. 
    The number of residual and skip channels was set to 64, and the convolution filter size was three. 
    The discriminator consisted of \minjaeedit{10} layers of non-causal dilated 1-D convolutions with leaky ReLU activation function ($\alpha = 0.2$). 
    The strides were set to \minjaeedit{1}, and linearly increasing dilations were applied to the 1-D convolutions, except the first and last layers, from \minjaeedit{1} to \minjaeedit{8}. 
    The number of channels and filter size were the same as the generator. 
    We applied weight normalization to all convolutional layers for both the generator and the discriminator \cite{salimans2016weight}.

    %%% TABLE: STFT
    \begin{table}[!t]
    \begin{center}         
    \caption{The details of the MR-STFT loss calculations. A Hanning window was applied before the FFT process.}
    \label{tab:stft}
    \scalebox{0.95}{
	{\small 
	\begin{tabular}{cccc}
	\Xhline{2\arrayrulewidth}
	STFT loss & FFT size & Window size & Frame shift \\
			\hline \hline
        $L_{\mathrm{stft}}^{(1)}$ & 512 & 240 (10 ms) & 50 ($\approx$ 2 ms) \\
        $L_{\mathrm{stft}}^{(2)}$ & 1024 & 600 (25 ms) & 120 (5 ms) \\
        $L_{\mathrm{stft}}^{(3)}$ & 2048 & 1200 (50 ms) & 240 (10 ms) \\
    \Xhline{2\arrayrulewidth}
	\end{tabular}} 
	}
	\end{center}         
	%\vspace*{-12pt}
	\end{table}
	%%% TABLE: STFT

    The MR-STFT loss was calculated by summing three STFT losses \reunwooedit{as shown in Table~\ref{tab:stft}, which had been defined in its original version} \cite{yamamoto2020parallel}.
    In the proposed method, \reunwooedit{
    to obtain the time-invariant masking filter in equation~(\ref{eq:az}), all the LSFs ($p=40$) collected from the training data were averaged, and converted to the corresponding LP coefficients \cite{soong1984line}.
    For a stable convergence, the masking filter's magnitude response was normalized to have a range from 0.5 to 1.0 before \rryuichiedit{applying \reunwooedit{it}} to the MR-STFT loss.
    }
    The discriminator loss was computed by the average of per-time-step scalar predictions with the discriminator.
    The value of hyperparameter, $\lambda_{\mathrm{adv}}$, in equation~(\ref{eq:gloss}) was chosen to be 4.0.
    \fryuichiedit{The} models were trained for 400 K steps with RAdam optimization ($\epsilon=1e^{-6}$) to stabilize training \cite{liu2019radam}.
    \rminjaeedit{T}he discriminator was fixed for the first 100 K steps, and both the generator and discriminator were jointly trained afterwards.
    The minibatch size was set to \minjaeedit{8}, and the length of each audio clip was set to 24 K time samples (1.0 second). 
    The initial learning rate was set to 0.0001 and 0.00005 for the generator and discriminator, respectively. 
    The learning rate was reduced by half every 200 K steps. 
    
    Across all vocoding models, the input auxiliary features were up-sampled by nearest neighbor up-sampling followed by 2-D convolutions so that the time-resolution of the auxiliary features matched the sampling rate of the speech waveforms \cite{odena2016deconvolution,Yamamoto2019}.
    
    \subsubsection{Text-to-speech generation}
    
    In the synthesis step, \reunwooedit{the acoustic} feature vectors were predicted by the acoustic model with the given input text. 
    To enhance spectral clarity, an LSF-sharpening filter was applied to the spectral parameters \cite{song2017effective}.
    By using these features as the conditional inputs, vocoding models such as WaveNet and Parallel WaveGAN generated corresponding time-sequences of the waveforms.

    \subsection{Evaluations}

    \reunwooedit{Naturalness} MOS tests were conducted to evaluate the perceptual quality of the proposed system\footnote{Generated audio samples are available at the following URL:\\ \url{https://sewplay.github.io/demos/wavegan-pwsl}}. 
    \minjaeedit{20} native Korean speakers were asked to make quality judgments about the synthesized speech samples using the five following possible responses: 1 = Bad; 2 = Poor; 3 = Fair; 4 = Good; and 5 = Excellent. 
    In total, 30 utterances were randomly selected from the test set and synthesized using the different generation models. 
    
    Table~\ref{table:mos} presents the MOS test results for the TTS systems with respect to the different vocoding models, and the analysis can be summarized as follows: First, in systems with autoregressive WaveNet vocoders, applying the noise-shaping filter performed significantly better than the plain systems (Tests 1 and 2).
    \eunwooedit{This} confirms that reducing auditory noise in the spectral valley regions was beneficial to perceptual quality. 
    However, the effectiveness of the noise-shaping filter was not evident for the Parallel WaveGAN systems (Tests 3 and 4). 
    Since the training and generation processes are both non-autoregressive, it \eunwooedit{might} be that the characteristics of a \ryuichiedit{noise-\eunwooedit{shaped}} target signal \eunwooedit{were} difficult for the model to capture without previous time-step information.
    Second, the systems with Parallel WaveGAN and the proposed perceptually weighted MR-STFT loss function demonstrated improved quality of synthesized speech (Tests 3 and 5). 
    %Because the weighting helped the model reduce generation errors in the spectral valleys, and because the adversarial training method helped capture the characteristics of realistic speech waveforms, the system was able to generate a natural voice within a non-autoregressive framework. 
    Because the weighting helped the model reduce generation errors in the spectral valleys, and because the adversarial training method helped capture the characteristics of realistic speech waveforms, the system was able to generate a natural voice within a non-autoregressive framework, \feunwooedit{providing the 14.87 K times faster inference than the best autoregressive model (Test2).}
    %\feunwooedit{Although its perceptual quality was slightly worse than the best autoregressive model (Test2), but its merits could be found in the fast inference speed, which was 294.12 times faster than the autoregressive one.}
    Consequently, the TTS system with the proposed Parallel WaveGAN vocoder achieved \reunwooedit{4.21 and 4.26} MOS results for female and male speakers, respectively.

    %%% TABLE: MOS TEST RESULTS
	\begin{table}[!t]   
	\begin{center}         
	\caption{
	\reunwooedit{Naturalness} MOS \reunwooedit{test} results with 95\% confidence intervals for the TTS systems with respect to the different vocoding models: The MOS results for the proposed system are in bold font. \reunwooedit{The KRF and KRM denote Korean female and male speakers, respectively.}}
	%\vspace*{-3pt}
	\label{table:mos}
	\scalebox{0.95}{
	{\small 
	\begin{tabular}{clcc}
	\Xhline{2\arrayrulewidth}
	Index & ~~~~~~~~~~~~~Model & KRF & KRM \\
			\hline \hline
	Test 1 & WaveNet & 3.64$\pm$0.14 & 3.60$\pm$0.13 \\
	Test 2 & WaveNet + NS & 4.36$\pm$0.11 & 4.32$\pm$0.10 \\
	Test 3 & Parallel WaveGAN & 4.02$\pm$0.10 & 4.11$\pm$0.11 \\%\hline
	Test 4 & Parallel WaveGAN + NS & 2.34$\pm$0.10 & 1.72$\pm$0.09  \\
	\textbf{Test 5} & \textbf{Parallel WaveGAN + PW} & \textbf{4.26$\pm$0.10} & \textbf{4.21$\pm$0.10} \\
	\hline
	Test 6  & Raw & 4.64$\pm$0.07 & 4.59$\pm$0.09 \\%\hline
    \Xhline{2\arrayrulewidth}
	\end{tabular}} 
	}
	\end{center}         
	%\vspace*{-12pt}
	\end{table}
    %%% TABLE: MOS TEST RESULTS           
    
\section{Conclusions}
    This paper proposed a spectral-domain perceptual weighting technique for Parallel WaveGAN-based TTS systems. 
    A frequency-dependent \eunwooedit{masking filter} was applied to the MR-STFT loss function, enabling the system to penalize errors near the spectral valleys. 
    As a result, the generation errors in those frequency regions were reduced\fryuichiedit{,  which} improved the quality of the synthesized speech.
    The experimental results verified that a TTS system with the proposed Parallel WaveGAN vocoder performs better than systems with conventional methods.
    \minjaeedit{Future research \fryuichiedit{includes} further improving the Parallel WaveGAN's perceptual quality by replacing the time-invariant spectral masking filter with a signal-dependent adaptive predictor.}

%\section*{Acknowledgment}
%\section{Acknowledgment}
%The authors would like to thank \minjaeedit{Hyeon-Kyeong Shin, \eunwooedit{Hyungseob Lim, Jihyun Lee}} and Suhyeon Oh at DSP\&AI Lab., Yonsei University, Seoul, Korea, for their support.

\bibliographystyle{IEEEtran.bst}
\bibliography{mybib}

\end{document}